\documentclass[superscriptaddress,aps,showpacs,nofootinbib,twocolumn]{revtex4}
\usepackage{graphicx}
\usepackage{amsmath,amssymb}

\def\CTP{{\it Commun. Theor. Phys.} }

\def\JHEP{{\it JHEP} }

\def\PL{{\it Phys. Lett.} }
\def\PR{{\it Phys. Rev.} }
\def\PRL{{\it Phys. Rev. Lett.} }

\def\frac#1#2{{\textstyle{{#1}\over {#2}}}}

\def\lsim{\mathrel{\rlap{\lower4pt\hbox{\hskip1pt$\sim$}}
    \raise1pt\hbox{$<$}}}
\def\gsim{\mathrel{\rlap{\lower4pt\hbox{\hskip1pt$\sim$}}
    \raise1pt\hbox{$>$}}}
\def\sqr#1#2{{\vcenter{\vbox{\hrule height.#2pt
         \hbox{\vrule width.#2pt height#1pt \kern#1pt
         \vrule width.#2pt}
         \hrule height.#2pt}}}}

\def\beq{\begin{equation}}
\def\eeq{\end{equation}}
\def\beqa{\begin{eqnarray}} 
\def\eeqa{\end{eqnarray}}

\def\laq{\raise 0.4 ex \hbox{$<$}\kern -0.8 em\lower 0.62 ex\hbox{$\sim$}}
\def\gaq{\raise 0.4 ex \hbox{$>$}\kern -0.7 em\lower 0.62 ex\hbox{$\sim$}}

\begin{document}

\title{Scaling of variables and the relation between noncommutative parameters in  Noncommutative Quantum Mechanics}

\author{O. Bertolami}
\altaffiliation{Email address: orfeu@cosmos.ist.utl.pt}

\author{J. G. Rosa}
\altaffiliation{Email address: joaopedrotgr@sapo.pt}
 
\affiliation{ Departamento de F\'\i sica, Instituto Superior T\'ecnico \\
Avenida Rovisco Pais 1, 1049-001 Lisboa, Portugal}

\author{C. M. L. de Arag\~ao}
\altaffiliation{Email address: cristiane.aragao@ct.infn.it}

\author{P. Castorina}
\altaffiliation{Email address: paolo.castorina@ct.infn.it}

\author{D. Zappal\`a}
\altaffiliation{Email address: dario.zappala@ct.infn.it}

\affiliation{Department of Physics, University of Catania and INFN-Sezione di Catania, Citt\'a Universitaria, Via S. Sofia 64, Catania, Italy}

\vskip 0.5cm

\date{\today}

\begin{abstract}

We consider Noncommutative Quantum Mechanics with phase space noncommutativity. In particular, we show that a scaling of variables leaves the noncommutative algebra invariant, so that only the self-consistent  effective parameters of the model are physically relevant.
 We also discuss the recently proposed relation of direct proportionality between the noncommutative parameters, showing that it has a limited applicability.

\end{abstract}

\pacs{11.30.Cp, 04.60.-m, 11.10.Ef \hspace{2cm}Preprint DF/IST-7.2005} 

\maketitle
 

\section{Introduction}

Recently there has been a growing interest  on the issue of noncommutative geometry. Since the discovery in string theory that the low-energy effective theory of a D-brane in the background of a NS-NS B field lives in a noncommutative space \cite{Connes, Seiberg}, many efforts have been devoted to the study of noncommutative field theories (NCFT's) and, in a non-relativistic approach, versions of noncommutative Quantum Mechanics (NCQM) \cite{Ho, Nair, Duval, Horvathy, 
Zhang_1, Zhang_2, Demetrian, Gamboa, Li, Acatrinei, Bertolami_1}.

Although in string theory only the coordinates space exhibits a noncommutative structure, some authors have studied models in which a noncommutative geometry defines the whole phase space \cite{Zhang_1,Bertolami_1,Djemai}. Noncommutativity between momenta arises naturally as a consequence of noncommutativity between coordinates, as momenta are defined to be the partial derivatives of the action with respect to the noncommutative coordinates \cite{Singh}. In a 4-dimensional space, this type of phase space structure is defined through the following algebra:
\beqa \label{noncommutation_1}
\lbrack x^{\mu},x^{\nu}]&=&i\theta^{\mu\nu}~,\qquad
\lbrack p^{\mu},p^{\nu}]=i\eta^{\mu\nu}~,\nonumber\\
\lbrack x^{\mu},p^{\nu}]&=&i\hbar\delta^{\mu\nu}~,
\eeqa
where the parameters $\theta^{\mu\nu}$ and $\eta^{\mu\nu}$ are antisymmetric. This algebra is consistent with usual Quantum Mechanics through the last commutation relation in the set of Eqs. (\ref{noncommutation_1}).

In a recent paper \cite{Bertolami_1}, we have considered a 2-dimensional version of this algebra, defined by the noncommutative parameters $\theta$ and $\eta$ and by Planck's constant $\hbar$ in the following way:
\beqa \label{noncommutation_2}
\lbrack x,y]&=&i\theta~,\qquad
\lbrack p_x,p_y]=i\eta~,\nonumber\\
\lbrack x_i,p_j]&=&i\hbar\delta_{ij}\qquad i=1,2~.
\eeqa

The implementation of this algebra can be done by constructing the noncommutative variables  $\{x,y,p_x,p_y\}$ from the commutative variables $\{x',y',p_x',p_y'\}$ by means of linear transformations. 

Two sets of linear transformations have been used
to obtain the complete algebra, Eqs. (\ref{noncommutation_2}) . 
The first kind of transformation only affects the variables $x$ and $p_y$, while the second one modifies the variables $y$ and $p_x$, thus leading to different results. To discuss  this ambiguity let us consider another type of linear transformations, which modifies simultaneously all variables:
\beqa \label{linear_1}
x&=&\xi\bigg(x'-{\theta\over2\hbar}p_y'\bigg)~,\qquad
y=\xi\bigg(y'+{\theta\over2\hbar}p_x'\bigg)~,\nonumber\\
p_x&=&\xi\bigg(p_x'+{\eta\over2\hbar}y'\bigg)~,\qquad
p_y=\xi\bigg(p_y'-{\eta\over2\hbar}x'\bigg)~,
\eeqa 
where $\xi$ is a scaling factor, which we have set to $1$.
In this way, the commutation relations between coordinates and the ones between momenta in Eq. (\ref{noncommutation_2}) are 
easily recovered, but the commutation relation between coordinates and momenta  is changed to:
\beq \label{noncommutation_3}
[x_i,p_j]=i\hbar_{eff}\delta_{ij}\qquad i=1,2~.
\eeq
where we have defined the \emph{effective Planck constant} $\hbar_{eff}=\hbar\big(1+\theta\eta/4\hbar^2\big)$.

In a 4-dimensional space, the generalization of the linear transformations Eq. (\ref{linear_1}) may be written as:
\beq \label{linear_2}
x^{\mu}=\xi\Bigg(x'^{\mu}-{\theta^{\mu}_{\ \nu}\over2\hbar}p'^{\nu}\Bigg)~,\qquad
p^{\mu}=\xi\Bigg(p'^{\mu}+{\eta^{\mu}_{\ \nu}\over2\hbar}x'^{\nu}\Bigg)~.
\eeq

By choosing $\xi=1$, these transformations lead to the following 4-dimensional algebra:
\beqa \label{noncommutation_4}
\lbrack x^{\mu},x^{\nu}]&=&i\theta^{\mu\nu}~,\qquad
\lbrack p^{\mu},p^{\nu}]=i\eta^{\mu\nu}~,\nonumber\\
\lbrack x^{\mu},p^{\nu}]&=&i\hbar\bigg(\delta^{\mu\nu}+{\theta^{\mu\alpha}\eta^{\nu}_{\ \alpha}\over4\hbar^2}\bigg)~.
\eeqa
Hence, one can define a \emph{4-dimensional effective Planck constant} as:
\beq \label{Planck_constant_1}
\hbar_{eff}=\hbar\bigg(1+{Tr[\theta\eta]\over4\hbar^2}\bigg)~.
\eeq 
and, moreover, it is clear that the commutation relation between coordinates and momenta is not diagonal, the off-diagonal elements being proportional to products of $\theta^{\mu\nu}$ and $\eta^{\mu\nu}$. 

We have used in Ref. \cite{Bertolami_1} the linear transformations Eqs. (\ref{linear_1}) to determine the noncommutative Hamiltonian for a particle moving in the $xy$ plane subject to a uniform gravitational field $\mathbf{g}=-g\mathbf{e_x}$. By considering
the effects of  phase space noncommutativity as small perturbations,
 we compared the leading order corrections to the commutative Hamiltonian with the experimental results obtained by Nesvizhevsky \emph{et al.} \cite{Nesvizhevsky} for neutrons in the Gravitational Quantum Well (GQW) formed by a horizontal mirror and the Earth's gravitational field. In this way, we have determined an upper bound of about meV/c for the fundamental momentum scale introduced by noncommutativity, $\sqrt{\eta}$. Assuming that the fundamental noncommutative length scale, $\sqrt{\theta}$, is smaller than the average neutron size, that is, 
of order $1~fm$, we have concluded that $(\hbar_{eff}-\hbar)/\hbar\lesssim O(10^{-24})$. Therefore, the difference between 
the effective Planck's constant and the usual Planck constant has no practical effects.

On the other hand, it has been recently claimed in Ref. \cite{Zhang_3} that for a particular choice of the 
value of the scaling factor $\xi$ the commutation relation between coordinates and momenta is exactly 
defined by the usual Planck constant.

In this paper, we argue that NCQM models with or without corrections to the value of Planck's constant are, in fact, physically equivalent, differing only in the way one defines the noncommutative parameters. 
We also analyze the direct proportionality between the noncommutative parameters proposed in \cite{Zhang_3}, emphasizing some of the problems that arise from its derivation and that affect its applicability.


\section{Scaling of Variables}

The role of the parameter $\xi$ in the linear transformations Eqs. (\ref{linear_1}) clearly corresponds to a simple scale transformation of the coordinates and of the momenta. This type of rescaling  only refers to the way through which an observer measures these variables. A  physical model cannot depend on the method  used to measure the dynamical quantities and, therefore, the considered noncommutative model must be invariant under scale transformations. This implies that a particular choice of the value of $\xi$ has no physical meaning.

To better appreciate this point, notice that, 
for an arbitrary value of $\xi$, one can always write the noncommutative algebra in the following way:
\beqa \label{noncommutation_5}
\lbrack x,y]&=&i\xi^2\theta\equiv i\theta_{eff}~,\qquad
\lbrack p_x,p_y]=i\xi^2\eta\equiv i\eta_{eff}~,\nonumber\\
\lbrack x_i,p_j]&=&i\xi^2\hbar\big(1+\frac{\theta\eta}{4\hbar^2}\big)\delta_{ij}\equiv i\hbar_{eff}\delta_{ij},\ i=1,2~.
\eeqa

The only experimentally measurable quantities of the model are, thus, $\theta_{eff}$, $\eta_{eff}$ and $\hbar_{eff}$. The choice of the value of $\xi$ only affects the way in which one writes these quantities in terms of the parameters $\theta$, $\eta$ and $\hbar$. 

For $\xi=(1+\theta\eta/4\hbar^2)^{-1/2}$, as proposed in Ref. \cite{Zhang_3}, one obtains:
\beq \label{NCparameters_1}
\theta_{eff}={\theta\over1+{\theta\eta\over4\hbar^2}},\  \eta_{eff}={\eta\over1+{\theta\eta\over4\hbar^2}}, \  \hbar_{eff}=\hbar ~.
\eeq

For $\xi=1$, as we have considered in Ref. \cite{Bertolami_1}, which means the scale transformation is included in the definition of the coordinates and the momenta, one obtains:
\beq \label{NCparameters_2}
\theta_{eff}=\theta,\  \eta_{eff}=\eta, \  \hbar_{eff}=\hbar\bigg(1+{\theta\eta\over4\hbar^2}\bigg) ~.
\eeq

Both models are physically equivalent, as one does not measure directly the values of $\theta$, $\eta$ or $\hbar$. As long as the definitions of $\theta_{eff}$, $\eta_{eff}$ and $\hbar_{eff}$ are consistent with each other, there is no reason to privilege any particular choice of $\xi$. One could think that maintaining the canonical Heisenberg commutation relation $[x_i,p_j]=i\hbar\delta_{ij}$ would be an important feature of the NCQM model, so to ensure the consistency with ordinary commutative Quantum Mechanics. However, if in fact, Nature exhibits a noncommutative phase space geometry, the Planck constant $\hbar$ will have no direct physical meaning by itself, being as important as parameters $\theta$ and $\eta$ in the definition of the experimentally relevant quantities.

As can be easily seen from Eqs. (\ref{noncommutation_5}), there is no value of $\xi$ for which $\theta_{eff}=\theta$, $\eta_{eff}=\eta$ and $\hbar_{eff}=\hbar$ simultaneously. Thus, one has to choose which of the commutation relations in the NCQM algebra exhibits a simple form in terms of $\theta$, $\eta$ and $\hbar$, all choices leading to the same results.

The important thing to bear in mind is that, when choosing $\xi=1$, $\hbar_{eff}$ does not correspond to Planck's constant in ordinary Quantum Mechanics. In the same way, when one chooses $\xi=(1+\theta\eta/4\hbar^2)^{-1/2}$, the fundamental scales of length and momentum introduced by noncommutativity are not simply $\sqrt{\theta}$ and $\sqrt{\eta}$, but depend on all three parameters $\theta$, $\eta$ and $\hbar$.

In a 4-dimensional space, one will have $\hbar_{eff}=\hbar$ if and only if $\xi=\big(1+{Tr[\theta\eta]\over4\hbar^2}\big)^{-1/2}$. However, the commutation relation $[x^{\mu},p^{\nu}]$ is non-diagonal for all values of $\xi$. Therefore, in 4 dimensions, phase space noncommutativity changes the canonical Heisenberg commutation relation between coordinates and momenta independently of any scale transformation.


\section{The relation between noncommutative parameters}

Another important issue in the noncommutative model is whether a relation between the noncommutative parameters can 
be derived.

Recently, it has been proposed \cite{Zhang_3} that a direct proportionality between $\theta$ and $\eta$ arises as a natural consequence of requiring the same Bose-Einstein bosonic algebra  in both commutative and noncommutative phase space models.
Indeed, if one constructs the bosonic creation and annihilation operators from the noncommutative variables as a generalization of one-particle Quantum Mechanics and demands that these operators satisfy the algebra:
\beqa \label{BE_1}
\lbrack a_1,a_1^+]&=&[a_2,a_2^+]=1~,\nonumber\\
\lbrack a_1,a_2]&=&[a_1^+,a_2^+]=0~,
\eeqa
then one finds the following general relation between the noncommutative parameters:
\beq \label{NC_relation_1}
\eta=K\theta~,
\eeq
where $K$ is the proportionality constant, the value of which may depend on the system under study. According to Ref. \cite{Zhang_3}, this constant may be determined from the fundamental parameters which characterize the system under study and from the constants $\hbar$ and $c$ through dimensional arguments, i.e. the relation between the fundamental parameters of noncommutativity is not universal (cf. below).

The bosonic creation and annihilation operators satisfy the same commutation relations as the corresponding operators of one-particle Quantum Mechanics in commutative phase space.
 However, it is not clear that the two algebras have any connection in noncommutative phase space.
 The bosonic algebra is determined by requiring the wave function of a set of bosonic particles to be symmetric. Although this condition leads to the same algebra of the 2-dimensional isotropic harmonic oscillator, this is just accidental. In this way, one is not retaining the Bose-Einstein statistics in the noncommutative model when constructing the relevant operators by generalizing one-particle Quantum Mechanics.

On a more fundamental level, it is still an open issue if the introduction of the deformed Moyal product in noncommutative quantum 
field theory is consistent with the algebra in Eq.(11) between the creation and annihilation operators. For instance, in Ref. 
\cite{Balachandran}, it has been shown that a Lorentz invariant interpretation of the noncommutative theory yields a deformation 
of that algebra that modifies the standard correspondence between spin and statistics and introduces violations of the Pauli principle.

Thus, we believe that there is no strong argument supporting a direct proportionality between the noncommutative parameters $\theta$ and $\eta$. Not only this relation is derived from an unjustified assumption, but its applicability is also quite limited. In fact,  the determination of the constant $K$ is straightforward for a 2-dimensional isotropic oscillator and  many systems can be approximated by a harmonic oscillator of this type, but this is not the general case. For instance, in the GQW considered earlier, particles move in the linear potential created by the Earth's gravitational field. Such a potential cannot be approximated by a harmonic oscillator. Hence, the only way of determining the value of $K$ through 
dimensional arguments. 

This method is not only unsatisfying, as it is not based on fundamental principles, but it is also intrinsically ambiguous.
 For the GQW, e.g., there are several ways to construct a constant $K$ with dimensions of $(mass/time)^2$. Considering the parameters of the system, that is the gravitational acceleration $g$ and the particle mass $m$, and the fundamental constants $\hbar$ and $c$, one may have:
\beq \label{constant_1}
K=\bigg({g^2\hbar\over c^4}\bigg)^2,\ K=\bigg({m^4g^2\over\hbar}\bigg)^{2/3}, \  K=\bigg({m^2 c^2\over\hbar}\bigg)^{2}~.
\eeq

The proportionality constant can also be built from Newton's constant $G$, as for instance $K=(m^6G^2/\hbar^3)^2$ or $K=(c^3/G)^{2}$. How does one decide  the correct value $K$? One may argue that the second possibility in Eq. (\ref{constant_1}), corresponding to the choice of Ref. \cite{Zhang_3}, is the most plausible one because it considers both parameters $m$ and $g$. But this is not a fundamental argument and there is no unique way of determining the value of $K$ for the GQW.

Another issue about relation Eq. (\ref{NC_relation_1}) is its dependence on the system under study. As  $\theta$ is determined, for instance, by string theory, with the  same value for all systems, the corresponding value of the momentum scale is system-dependent.
 This means, for example, that  there is no absolute uncertainty principle for measuring momenta in orthogonal directions, 
which is, in our opinion, another unsatisfying feature of relation Eq. (\ref{NC_relation_1}). A relation such as Eq. (\ref{NC_relation_1}) does not reflect the fact that NCQM is, at best, an effective model of a more encompassing underlying theory. Therefore, it is unlikely that relationships between the parameters of the noncommutative algebra can be derived from the dynamics of any particular system.

Hence, we believe that there is no compelling reason supporting relation Eq. (\ref{NC_relation_1}). Although the algebra itself establishes relations between the noncommutative parameters which seem universal, it does not provide a way of determining the value of each of the three effective parameters from the value of the other two. For example, in the considered 2-dimensional model, one may write the following relations:
\beqa \label{NC_relation_2}
\theta_{eff}&=&\xi^2\bigg({\hbar_{eff}\over\hbar}-1\bigg){4\hbar^2\over\eta_{eff}}~,\nonumber\\
\eta_{eff}&=&\xi^2\bigg({\hbar_{eff}\over\hbar}-1\bigg){4\hbar^2\over\theta_{eff}}~,\nonumber\\
\hbar_{eff}&=&\xi^2\hbar\bigg(1+{1\over\xi^4}{\theta_{eff}\eta_{eff}\over4\hbar^2}\bigg)~,
\eeqa
which clearly illustrate the dependence on the values of the scaling factor and of the Planck constant, so that only a self consistent determination of the effective parameters is possible. 
If fundamental relations between the noncommutative parameters do, in fact, exist, 
they must be established by the underlying quantum field theory which gives origin to the 
noncommutative phase space geometry at the level of Quantum Mechanics.


\section{Conclusions}

In this paper we have analyzed some general features of  NCQM models where noncommutativity in both coordinates and momenta space is considered. 
In particular, we have shown that the noncommutative algebra is invariant under scale transformations, so that all values of the scaling factor $\xi$ describe the same physical model. Different choices of the value of this scale parameter lead to different definitions of $\theta_{eff}$, $\eta_{eff}$ and $\hbar_{eff}$. However the latter are quantities that can only be determined experimentally and  the parameters $\theta$, $\eta$ and $\hbar$ are relevant only to the extent that they establish relationships between the corresponding effective quantities.

Thus, one can consider a correction to Planck's constant only in the sense that the effective Planck constant may formally differ from the usual Planck constant $\hbar$ (for $\xi=1$), although the latter does not play a fundamental role on the NCQM model. The same arguments apply for the corrections to the noncommutative parameters $\theta$ and $\eta$ which arise from other values of $\xi$. Despite these formal considerations, such corrections are negligible, as we have shown in Ref. \cite{Bertolami_1}.

Although in a 2-dimensional space the commutation relation between coordinates and momenta is formally identical in both the commutative and the noncommutative models, this is not the case in 4-dimensions as this relation becomes non-diagonal when phase space noncommutativity is considered.

We also conclude that the recently proposed relation of direct proportionality between the parameters $\theta$ and $\eta$ 
is based on the unjustified assumption that the Bose-Einstein algebra is a generalization of one-particle Quantum Mechanics 
in noncommutative phase space. Furthermore, the proportionality constant one obtains is not system-independent, as one 
would expected for a NCQM model. Therefore, it is in our view pointless aiming to derive a universal relationship 
between the noncommutative parameters from the dynamics of a particular system and the use of 
dimensional arguments is ill-defined at best. 

In summary, we can say that the current state of our knowledge on the nature and origin of NCQM, does not allow the 
determination the three parameters that characterize the NCQM model from the values of only two of them. 
Actually, it is not clear at all whether such fundamental relationships do exist. Hopefully, future advancements 
in the search for the unified quantum field theory will bring new insights on this issue.

\vfill






\end{document}